\documentclass[a4paper,11pt]{article}
\usepackage[round,authoryear]{natbib}
\usepackage[german,american]{babel}
\UseRawInputEncoding
\usepackage[a4paper,top=3.5cm,bottom=3.5cm,left=3.0cm,right=3.0cm,marginparwidth=1.75cm]{geometry}
\usepackage{amsmath}
\usepackage{amsfonts}
\usepackage{amssymb}
\usepackage{graphicx}
\usepackage[font=small,labelfont=bf]{caption}
\usepackage{subcaption}
\usepackage{float}
\usepackage[OMLmathsfit]{isomath}
\usepackage{amssymb} 
\usepackage{amsxtra} 
\usepackage{bm} 
\usepackage{mathtools}

\usepackage[usenames]{color}
\definecolor{darkgreen}{rgb}{0.0,0.4,0.0}

\usepackage{verbatim}
\usepackage[colorlinks=true, allcolors=blue]{hyperref}
\usepackage{soul}
\usepackage{cleveref}

\DeclareCaptionLabelFormat{mylabel}{#1 #2\hspace{1.5ex}}
\usepackage{bbold}

\usepackage{hyphenat}
\newcommand{\secref}[1]{Sec.~(\ref{#1})}

\providecommand{\keywords}[1]
{
  {
  \small	
  \textbf{\textit{Keywords---}} #1
}}

\title{\textbf{An artificial neural network for surrogate modeling of stress fields in viscoplastic polycrystalline materials}}

\author{Mohammad S. Khorrami$^{1,*}$, Jaber R. Mianroodi$^{1,*}$, Nima H. Siboni$^{1}$, \\
        Pawan Goyal$^{2}$, Bob Svendsen$^{1,3}$, Peter Benner$^{2}$, Dierk Raabe$^{1}$ \\
       {\footnotesize $^1$Microstructure Physics and Alloy Design,} \\ 
        {\footnotesize Max-Planck-Institut f\"ur Eisenforschung, D\"usseldorf, Germany} \\
        {\footnotesize $^2$Computational Methods in Systems and Control Theory,} \\ 
        {\footnotesize Max Planck Institute for Dynamics of Complex Technical Systems, Magdeburg, Germany} \\
        {\footnotesize $^3$Material Mechanics, RWTH Aachen University, Aachen, Germany}\\
       {\footnotesize $^*$Corresponding Authors: m.khorrami@mpie.de, j.mianroodi@mpie.de}
} 

\begin{document}

\maketitle

\begin{abstract}
The purpose of this work is the development of an artificial neural network (ANN) for surrogate modeling of the mechanical response of viscoplastic grain microstructures. To this end, a U-Net-based convolutional neural network (CNN) is trained to account for the history dependence of the material behavior. The training data take the form of numerical simulation results for the von Mises stress field under quasi-static tensile loading. The trained CNN (tCNN) can accurately reproduce both the average response as well as the local von Mises stress field. The tCNN calculates the von Mises stress field of grain microstructures not included in the training dataset about 500 times faster than its calculation based on the numerical solution with a spectral solver of the corresponding initial-boundary-value problem. The tCNN is also successfully applied to other types of microstructure morphologies (e.g., matrix-inclusion type topologies) and loading levels not contained in the training dataset.
\end{abstract}

\keywords{Machine learning, Deep learning, Convolutional neural network, U-Net, Plasticity, Mechanical response, Spectral method}

\section{Introduction}
\label{sec:intro}
In material science and engineering, the modeling of the microstructural- and global mechanical response of heterogeneous polycrystalline and polyphase materials is important for understanding the underlying mechanisms behind their load-bearing and damage-related properties. Modeling such materials under consideration of their typically inelastic and nonlinear constitutive response and with realistic grain microstructures based on the numerical solution of corresponding initial-boundary-value problems (IVBPs) via spectral methods \cite[e.g.,]{shanthraj2015numerically, willot2015fourier, schneider2017fft, lucarini2019dbfft, roters2019damask, lebensohn2020spectral, khorrami2020development} is computationally expensive, especially at high resolution \cite[e.g.,][]{roters2010overview, roters2012damask, diehl2017identifying}. Among the possibilities for improvement of computational efficiency in material modeling, perhaps the most prominent ones in recent years, are approaches based on artificial neural networks (ANNs) and machine learning (ML), \cite[e.g.,][]{stoll2021machine}. Except for a few early works \cite[e.g.,][]{wu1991neural,haj2008nonlinear}, most of these approaches have been introduced in the last four years, \cite[e.g.,][]{ali2019application,mayer2021dislocation,pandya2020strain,settgast2020hybrid,mianroodi_lossless_2022}. Development of such ANN-based "surrogate" models involves the determination of network parameters via (constrained) optimization, i.e., so-called network "training". The data employed for this purpose can be experimental, empirical, or synthetic. An example of the latter results from the numerical solution of IVBPs based on physical models. In the "data-driven" case, training is based on such data alone. Going beyond this, one can employ physical relations (e.g., mechanical equilibrium) used to obtain the data as additional constraints, resulting in physics-constrained training. Note that this is different in character from the so-called physics-informed neural networks (PINN) \cite[e.g.,][]{raissi2019physics}. In this latter case, the data in question take the form of prescribed initial and boundary conditions of an IBVP, and the ANN is used to approximate the solution of the corresponding partial differential equation(s). The IBVP is then solved numerically via constrained optimization \cite[e.g.,][]{lagaris1998artificial}. A recent review of applications in the field of continuum mechanics and material modeling has been provided by \cite{bock2019review}. In the current work, attention is focused on the data-driven approach. The data are obtained from the numerical solution of a BVP for quasi-static mechanical equilibrium based on viscoplastic material modeling of grains in a grain microstructure.

A number of data-driven approaches have been proposed in solid mechanics. For example, \cite{yang2021deep} trained a conditional generative adversarial network (cGAN) to reproduce stress- and strain fields in mechanically loaded heterogeneous microstructures. \cite{mianroodi2021teaching} trained a U-Net-based ANN using results from the numerical simulations of mechanically stressed grain-scale microstructures for elastic and ideal viscoplastic constitutive single-grain response as well as different grain configurations. The resulting trained ANN produced high fidelity predictions of the von Mises stress field in the grain microstructure due to external loading. 

In the current work, a U-Net-based ANN is trained to calculate the evolution of the von Mises stress field in heterogeneous microstructures consisting of grains with elastic-viscoplastic constitutive response subject to uniaxial tensile loading. To this end, the ideal elastoplastic case originally considered by \cite{mianroodi2021teaching} is generalized here to viscoplasticity with linear isotropic strain hardening. In addition to the prediction of the stress field, the results are compared for the average von Mises stress of the grain microstructure versus the applied strain. The work begins in Section \ref{sec:methodology} with a brief description of the numerical solution of the BVP for mechanical equilibrium based on isotropic $J_{2}$ plasticity at finite deformation used to obtain synthetic training data. Section \ref{sec:setup} deals with dataset generation, the U-Net architecture, and training. The dependence of the trained convolutional neural network (tCNN) on details of the training dataset such as (i) the number of grains in the microstructure, and (ii) the range of material properties chosen for each grain, is discussed in Section \ref{sec:results}. In addition, the performance of the tCNN for (i) microstructure morphologies (e.g., matrix-inclusion), and (ii) loading levels, that had not been included in the training dataset is also studied and quantitatively evaluated. The work ends with conclusions and an outlook in Section \ref{sec:conclusions}.

\section{A brief review of J2 viscoplasticity with isotropic hardening}
\label{sec:methodology}
The data for the training of the ANN is obtained from the spectral solver for the BVP of the quasi-static mechanical equilibrium based on viscoplastic material modeling of grains in a grain microstructure. The purpose of this section is a brief review of the basic relations involved. In the current context of isothermal and quasi-static conditions, these include in particular mechanical equilibrium,
\begin{equation}
    \mathop{\mathrm{div}}\mathbf{P}=\bm{0},
\end{equation}
in terms of the first Piola-Kirchhoff stress $\mathbf{P}$. 
In the context of the viscoplasticity, the deformation gradient $\mathbf{F}$ is decomposed as 
\begin{equation}
    \mathbf{F}=\mathbf{F}_{\mathrm{e}} \mathbf{F}_{\mathrm{p}},
\end{equation}
into elastic $\mathbf{F}_{\mathrm{e}}$ and plastic $\mathbf{F}_{\mathrm{p}}$ deformation gradients.
The linear elastic relation 
\begin{equation}
    \mathbf{S}_{\mathrm{e}}=\lambda\,(\mathbf{I}\cdot\mathbf{E}_{\mathrm{e}})+2\mu\,\mathbf{E}_{\mathrm{e}}\cdot\mathbf{E}_{\mathrm{e}}
\end{equation}
is assumed between the elastic second Piola-Kirchhoff stress \(\mathbf{S}_{\mathrm{e}}\) and the elastic Green strain \(\mathbf{E}_{\mathrm{e}}=\frac{1}{2}(\mathbf{F}_{e}^{T} \mathbf{F}_{e}-\mathbf{I})\). In the following, the Lam\'{e} constants with \(\lambda=E\nu/((1+\nu)(1-2\nu))\) and \(\mu=E/(2(1+\nu))\) are determined in terms of the Young's modulus \(E\) and the Poisson ratio \(\nu\). Note that \(\mathbf{P}=\mathbf{F}_{\mathrm{e}}\mathbf{S}_{\mathrm{e}}\mathbf{F}_{\smash{\mathrm{p}}}^{-\mathrm{T}}\). 
The evolution of \(\mathbf{F}_{\mathrm{p}}\) is determined by the \(J_{2}\) flow rule
\begin{equation}
    \dot{\mathbf{F}}_\mathrm{p}\mathbf{F}_{\smash{\mathrm{p}}}^{-1}=\dot{\gamma}_{\mathrm{p}}\,\mathbf{S}_{\mathrm{e}}^{\mathrm{dev}}/|\mathbf{S}_{\mathrm{e}}^{\mathrm{dev}}|,
\end{equation}
where \(\dot{\gamma}_{\mathrm{p}}\) is the rate of equivalent plastic shear, and $\mathbf{S}_{\mathrm{e}}^{\mathrm{dev}}$ is the deviatoric part of \(\mathbf{S}_{\mathrm{e}}\). The viscoplastic (i.e., rate-dependent) form
\begin{equation}
    \dot{\gamma}_{\mathrm{p}}=\dot{\gamma}_0(|\mathbf{S}_{\mathrm{e}}^{\mathrm{dev}}|/\xi)^{n_{0}}
\end{equation}
for the evolution of \(\gamma_{\mathrm{p}}\) is determined by the typical material inelastic shear rate \(\dot{\gamma}_0\), the power-law exponent \(n_{0}\), and the flow resistance
\begin{equation}
    \xi(\gamma_{\mathrm{p}})=\xi_{0}+h_{0}\gamma_{\mathrm{p}}
\end{equation}
for linear isotropic hardening, with \(\xi_{0}\) the initial flow resistance, and \(h_{0}\) the isotropic hardening modulus. 
Material parameters of the viscoplastic model include $E$, $\nu$, \(\dot{\gamma}_0\), \(\xi_{0}\), \(h_{0}\), and \(n_{0}\). For network training, \(\dot{\gamma}_0\) and \(n_{0}\) are assumed to be the same for all grains in the grain microstructure. On the other hand, $E$, $\nu$, \(\xi_{0}\), and \(h_{0}\) are assumed to vary from grain to grain. The range of these material properties as well as the way of the selection of these values are presented in section \secref{Sec:training_dataset}. The values \(\dot{\gamma}_0=0.001\) and \(n_{0}=20\) are considered in this viscoplastic model.

\section{Neural network employed and training}
\label{sec:setup}
\subsection{Network input, architecture, and output}
The viscoplastic model has been implemented in the simulation software toolkit DAMASK \cite[][]{roters2019damask} using the spectral method. Training data is generated for the case of uniaxial tension via mean deformation gradient control under periodic boundary conditions in all directions. The network output is the scalar von Mises stress field \(\sigma_{\mathrm{vM}}=\sqrt{3\mathbf{T}_{\mathrm{dev}}\cdot\mathbf{T}_{\mathrm{dev}}/2}\) in terms of the Cauchy stress \(\mathbf{T}=\mathbf{P}\mathbf{F}^{\mathrm{T}}/\det\mathbf{F}\). The network can be applied to train all stress components; however, for simplicity, only von Mises stress is investigated in this paper.

\begin{figure}[H]
	\centering
	\includegraphics[width=0.6\textwidth]{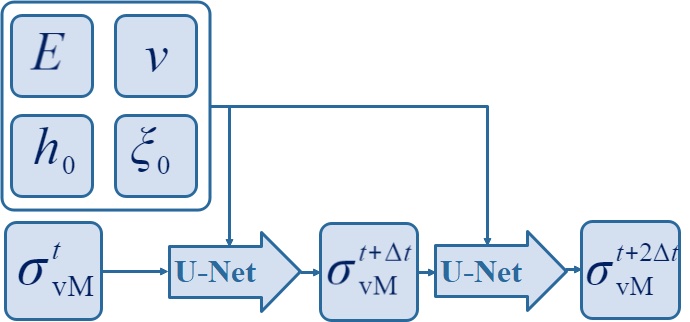}
	\includegraphics[width=0.39\textwidth]{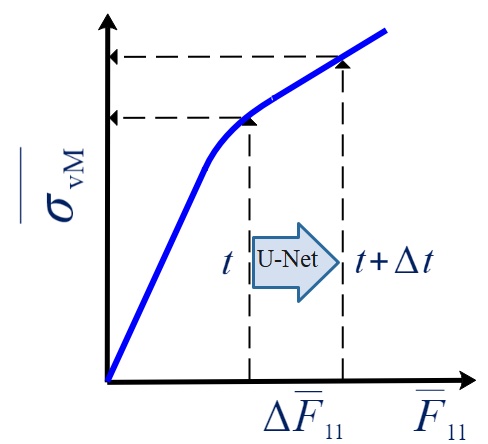}
	\caption{The schematic illustration of the machine-learning-based model for predicting the history-dependent local von Mises stress. $E$: Young's modulus, $\nu$: Poisson ratio, $\xi_0$: initial flow resistance, $h_0$: initial isotropic hardening, $\sigma_{\mathrm{vM}}$: von Mises stress, $\bar{F}_{11}$: normal component of mean deformation gradient.}
	\label{fig:diagram_ML}
\end{figure}

To obtain the von Mises stress field at load step $t+\Delta t$ (i.e., $\sigma_{\mathrm{vM}}^{t+\Delta t}$), material property distribution ($E, \nu, h_{0}, \xi_{0}$), as well as the von Mises stress field at the load step $t$ (i.e. $\sigma_{\mathrm{vM}}^{t}$), are required. Hence, these five fields, stored as images, are passed into the network as the input, and the output of the network is the next von Mises stress field (i.e. $\sigma_{\mathrm{vM}}^{t+\Delta t}$), see Figure \ref{fig:diagram_ML}. The network is obtained just for one increment-specific loading, for example, uniaxial extension. It should be noted that the four images (storing the distribution of $E, \nu, h, \xi_{0}$) are constant during the loading, while $\sigma_{\mathrm{vM}}$ is evolving. 

Since the network is dependent on the current values of $\sigma_{\mathrm{vM}}$, the model is history-dependent and able to capture viscoplastic mechanical behavior. The current ANN is based on the U-Net convolutional neural network architecture introduced by \cite{ronneberger2015u}. As it is shown by \cite{mianroodi2021teaching}, this architecture is suitable for surrogate modeling the stress field in solid mechanics problems. 
For simplicity, $\sigma_{\mathrm{vM}}$, is employed as the input and output of the ANN to train the network in a recursive manner.

\subsection{Training dataset}
\label{Sec:training_dataset}
\begin{figure}[H]
	\centering
	\includegraphics[width=0.24\textwidth]{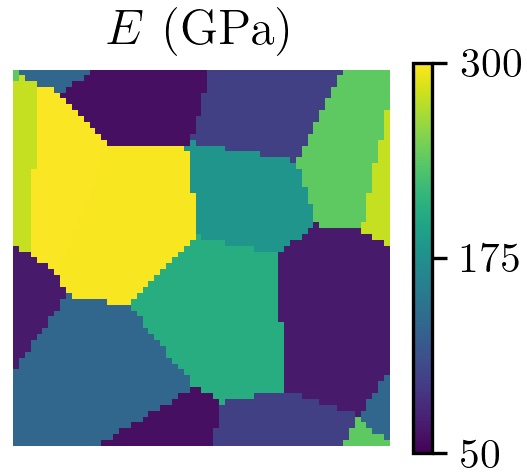}
	\includegraphics[width=0.24\textwidth]{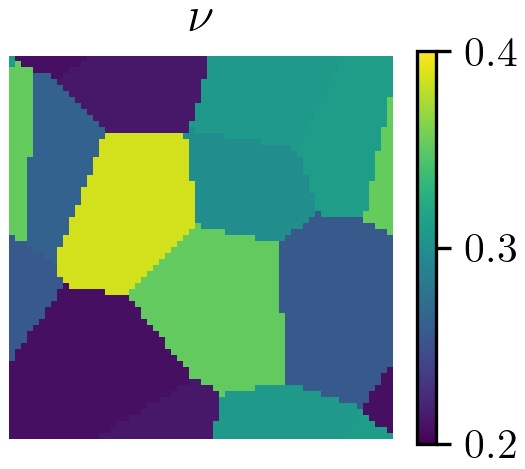}
	\includegraphics[width=0.24\textwidth]{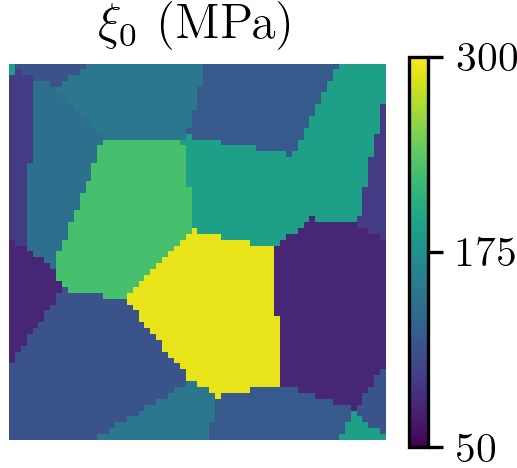}
	\includegraphics[width=0.24\textwidth]{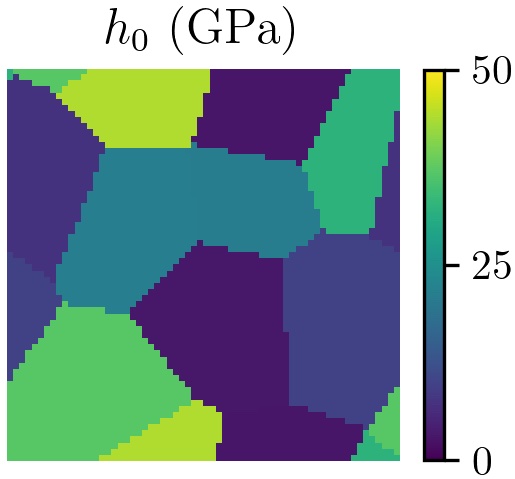}
	\caption{An example of a randomly generated microstructure and the material property distribution is shown. $E$: Young's modulus, $\nu$: Poisson ratio, $\xi_0$: initial flow resistance, $h_0$: initial isotropic hardening.}
	\label{fig:properties_dataset1}
\end{figure}

The dataset is obtained by generating random configurations of grains using Voronoi tessellation. Randomly selected material properties, i.e. elastic ($E$, $\nu$) and plastic ($\xi_0$, $h_0$) properties, are assigned to each grain. The open-source simulation package DAMASK \cite[][]{roters2019damask} is used to solve the mechanical BVPs via the spectral method in the J2-plasticity framework. The material properties are chosen randomly from a uniform distribution using minima and maxima values as shown in \Cref{tab:range_properties}. One specimen of the dataset is shown in Fig. \ref{fig:properties_dataset1}. 

\begin{table}[h]
	\centering
	\caption{The range of the material properties used for generating the dataset. $E$: Young's modulus, $\nu$: Poisson ratio, $\xi_0$: initial flow resistance, $h_0$: initial isotropic hardening.}
	\label{tab:range_properties}
	\begin{tabular}{|c|c|c|c|c|}
	    \hline
		& $E$ (GPa)  & $\nu$ & $\xi_0$ (MPa) & $h_0$ (GPa) \\ \hline
		Minimum & 50 & 0.2 & 50 & 0 \\
		Maximum & 300 & 0.4 & 300 & 50 \\
		\hline
	\end{tabular}
\end{table}

\begin{figure}[H]
    \centering
    \includegraphics[width=0.4\textwidth]{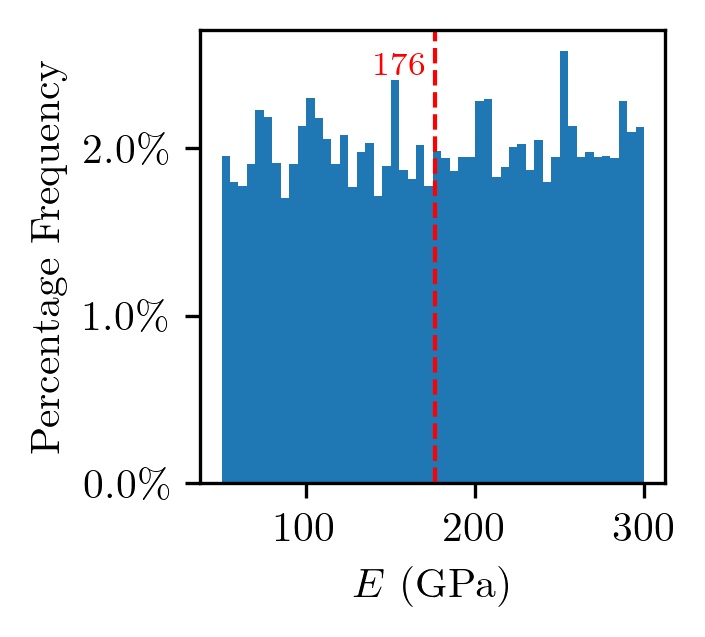}
    \includegraphics[width=0.4\textwidth]{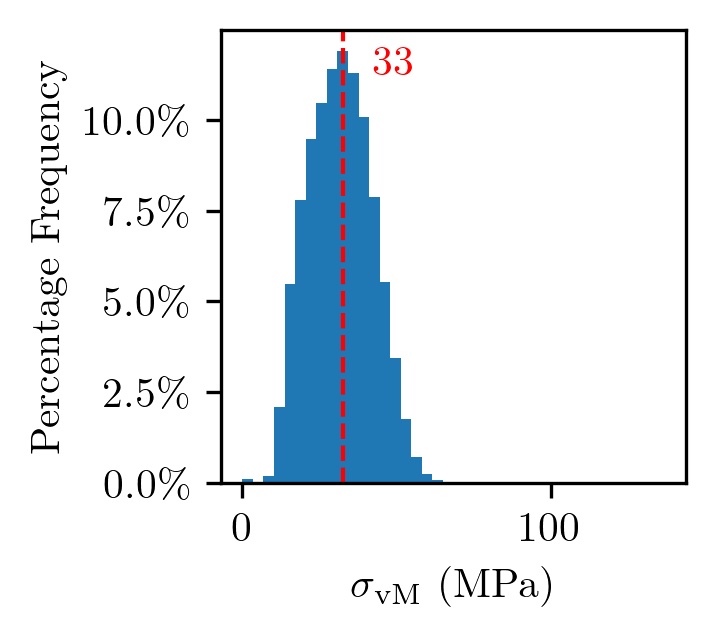}
    \includegraphics[width=0.4\textwidth]{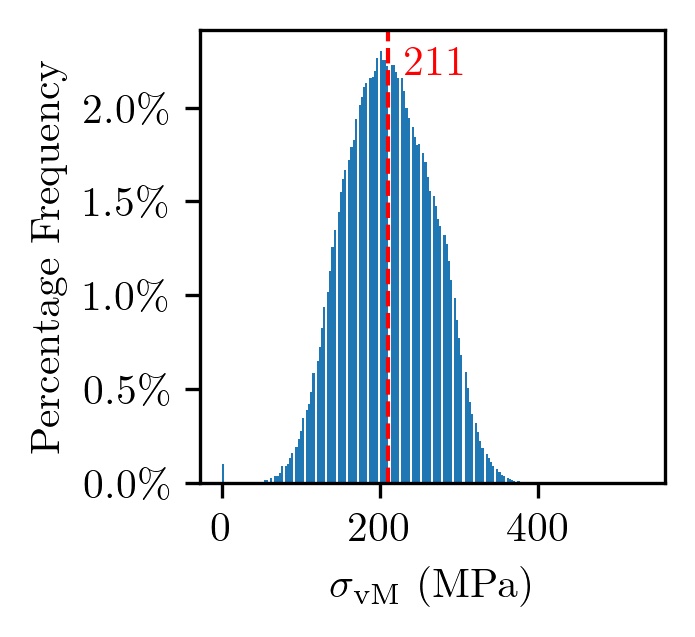}
    \includegraphics[width=0.4\textwidth]{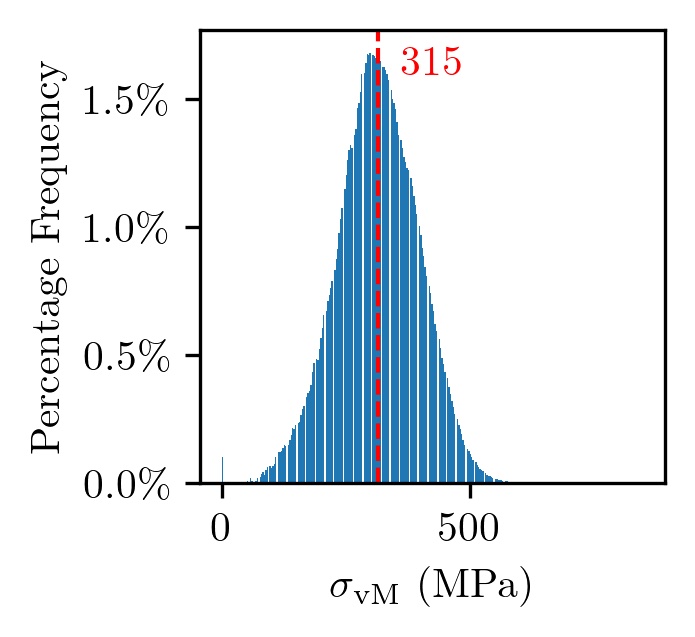}
    \caption{The histogram profile of $E$ for all grid points and all the configurations in the training dataset (top left), the distribution of $\sigma_{\mathrm{vM}}$ for all points at the increment of $t=1$ (top right) and $t=10$ (bottom left), and $t=20$ (bottom right) are shown.}
    \label{fig:hist}
\end{figure}

In this work, we generate 1000 various geometries each containing 10 grains. These are considered for the viscoplastic analysis under tensile loading. At each increment of the loading, the von Mises stress field is stored. We apply the uniaxial tensile loading through the average deformation gradient of the box in 40 increments with the rate of $0.001$ in the quasi-static condition. The first 20 increments are considered for the training data, and the last 20 increments are employed to test the ML model for extrapolating the loading. In Fig. \ref{fig:hist}, the histogram of the distribution of the Young's modulus is plotted for all the points in the whole dataset. The other material properties, e.g. $\nu$, $\xi_0$, and $h_0$, also follow a uniform distribution.

Fig. \ref{fig:hist} also depicts the histogram profile of the $\sigma_{\mathrm{vM}}$ at load step $(t=1)$, ($t=10$), and $(t=20)$ with the mean values of 33, 211, and 315 MPa, respectively. These histogram profiles are plotted for all geometries and pixels at a certain increment of the loading in the whole dataset. According to this figure, the distribution of $\sigma_{\textrm{vM}}$ at each step of the loading is similar to a Gaussian shape with the specific mean values mentioned above, which corresponds to the level of the loading.

\subsection{The deep neural network architecture and the training process}
\begin{figure}[H]
	\includegraphics[width=1\textwidth]{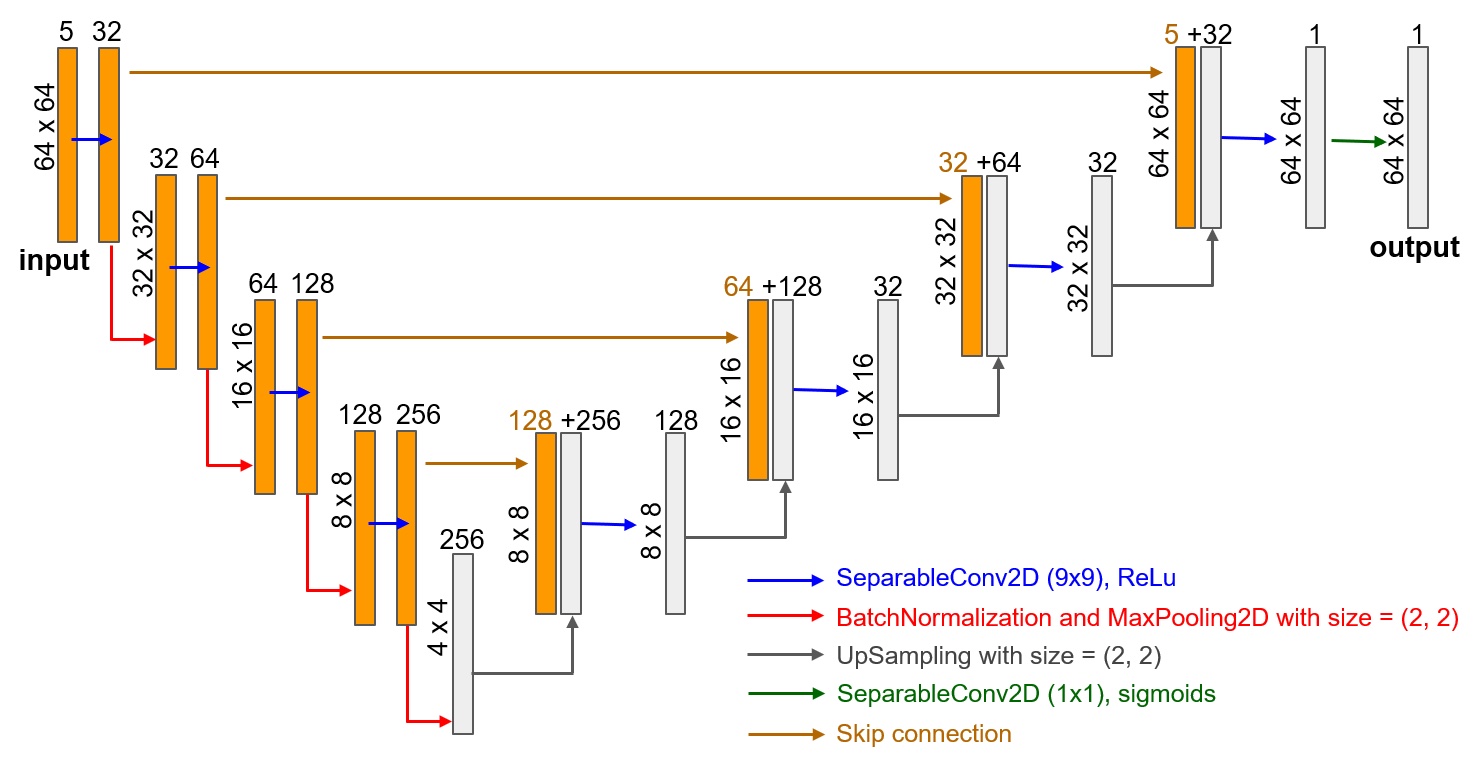}
    \caption{The architecture of the U-Net used in this work consists of two-dimensional (2D) separable convolution with the ReLU activation function, batch normalization, 2D max pooling, and upsampling with bilinear interpolation. Note that the standard terminology of TensorFlow \cite[][]{tensorflow2015-whitepaper} for these operations is used.}
    \label{fig:unet}
\end{figure}

The neural network used in this work is similar to the network presented by \cite{mianroodi2021teaching}. Fig. \ref{fig:unet} schematically depicts the network architecture, which is called U-Net \cite[][]{ronneberger2015u} due to its shape. In this work, the size of the input and output images is $64$ by $64$. Since $E$, $\nu$, $h_0$, $\xi_0$, and the current value for $\sigma_{\mathrm{vM}}$ serve as input images, we have 5 input channels. In addition, we consider 32 filters to capture the main features from the input images. As shown in the figure, four types of operation are performed in the U-Net, namely separable two-dimensional (2D) convolution with a kernel size of $9\time9$, batch normalization, 2D max pooling, and 2D upsampling with bilinear interpolation. For the training process, we consider Adam optimizer with a learning rate of $0.001$. The loss function is the mean absolute error (MAE) between the predicted von Mises stress and the one calculated with DAMASK. We used TensorFlow \cite[][]{tensorflow2015-whitepaper} to create and train the network. The ML model is trained for 500 epochs with the final MAE of $1.733$ MPa and $1.743$ MPa for the training and test dataset, respectively. It should be noted that the ML model is obtained without any sign of overfitting.

\section{Results and Discussion}
\label{sec:results}
In this section, we discuss and compare the results of the spectral method and the ML-based calculation of the stress field as well as the mechanical response of polycrystalline materials. In this regard, among all simulation data and predicted $\sigma_{\mathrm{vM}}$, some of them are presented here to analyze the quality of the predictions. Firstly, we investigate the ML-based prediction of the stress field for geometries similar to the dataset. In this part, we examine cases similar to the training dataset but with different numbers of grains, as well as the cases with high contrast material properties in the neighboring grains. Secondly, we explore the ML model for geometries far from the dataset, e.g., composite microstructures or matrix-inclusion configurations. Thirdly, we examine the ML model in extrapolating the loading beyond the one used for the training dataset. Finally, the ML model and the spectral method are compared in terms of the computational time and efficiency in predicting the stress fields. 

\subsection{Model microstructures with varying numbers of grains}
Here, we examine three types of simulation boxes containing 5, 10, and 20 grains. Although the training dataset only includes simulation boxes with 10 grains, the ML model can be employed for other grain sizes as well. To test the trained model, 50 test cases for each configuration of 5 grains, 10 grains, and 20 grains (a total of 150 cases) are randomly generated and analyzed. As shown in Table \ref{tab:grains}, the mean average error in these test datasets is quite low, close to the MAE of the training dataset. In this comparison, the MAE of the case for 5-grain polycrystals is below that for the cases of 10- and 20-grain-containing microstructures. The main reason for this is associated with the length of the grain boundaries. It can be noticed that the regions of largest errors are mostly occurring around the boundaries. Therefore, a plausible reason for the grain size dependence of the error is that more grains result in a larger total length of the grain boundaries, thus, larger averaged errors. In what follows, we discuss the stress and the error fields for these three cases separately.

\begin{table}[h]
		\centering
		\label{tab:1}
		\caption{The mean absolute error (MAE) of the test dataset for geometries with different numbers of grains. Note that the von Mises stress in the dataset can reach values above 500 MPa, as shown in Fig. \ref{fig:hist} bottom right.}
		\begin{tabular}{|c|c|c|c|c|}
			\hline
			number of grains & 5  & 10 & 20 \\ \hline
			MAE (MPa) & 1.578 & 1.678 & 1.869 \\
		\hline
		\end{tabular}
		\label{tab:grains}
\end{table}

\begin{figure}[H]
    \centering
    \includegraphics[width=1\textwidth]{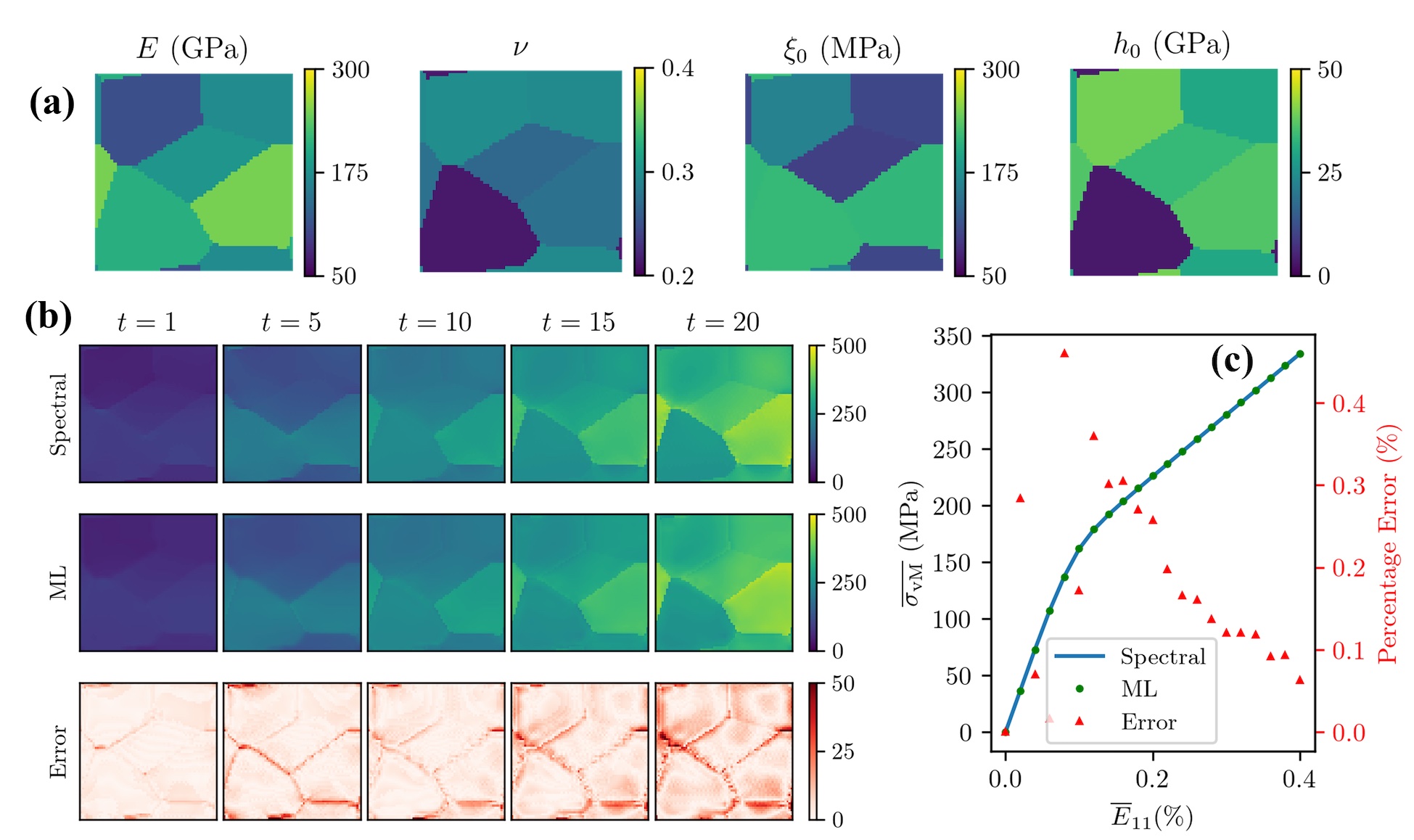}
    \caption{\textbf{(a)} Material properties for a microstructure patch consisting of 5 grains. \textbf{(b)} The local von Mises stress fields based on the solution obtained by the spectral method (above) and the ML model (middle). Also, the absolute error between the ML and the spectral method is plotted (bottom), Units: MPa. \textbf{(c)} The mean von Mises stress and the error (in percentage) between the ML model and the spectral method vs. the strain, the normal component of the mean Green-Lagrange strain, $\bar{E}_{11}$.}
    \label{fig:5grains}
\end{figure}

We show that the ML model can predict the mechanical response of polycrystals containing larger and smaller grain sizes compared to the ones used for the ML training. The first example is investigated for a 5-grain polycrystal, which has not been a part of the training process. The chosen material properties of this example are shown in Fig. \ref{fig:5grains} (a). The contours of $\sigma_{\mathrm{vM}}$ obtained by using the spectral method and the ML model are shown at different load steps in Fig. \ref{fig:5grains} (b). As can be noted in the figure, the ML-predicted von Mises stress is very close to the spectral solution. In this figure, the absolute error between the results of the ML model and the spectral method is also depicted. It should be noted that the absolute error increases at higher deformation since the error is accumulated in the next steps of the loading as shown in Fig. \ref{fig:5grains} (b). However, the maximum error of each pixel is less than $10 \%$, or $50$ MPa. The mean von Mises stress versus the mean normal component of the Green-Lagrange strain, $\bar{E}_{11}$, is plotted based on the spectral method and the ML model in Fig. \ref{fig:5grains} (c). In this curve, the relative error in percentage is also plotted, which is defined as follows:
\begin{equation}
    \mathrm{Percentage~Error} = \frac{\overline{\sigma_{\mathrm{vM}}^{\mathrm{ML}}} - \overline{\sigma_{\mathrm{vM}}^{\mathrm{Spectral}} }}{ \overline{\sigma_{\mathrm{vM}}^{\mathrm{Spectral}}}} \times 100,
\end{equation}
where $\overline{\sigma_{\mathrm{vM}}^{\mathrm{ML}}}$ and $\overline{\sigma_{\mathrm{vM}}^{\mathrm{Spectral}}}$ are the current values of the mean von Mises stress for the whole simulation box. According to the figure, the prediction of the ML model for the mean value of the von Mises stress is very close to the results of the spectral solver with a maximum error of less than $0.5\%$. It should be noted that the relative errors during the very first deformation steps are larger than other steps due to the smaller value of $\overline{\sigma_{\mathrm{vM}}^{\mathrm{Spectral}}}$.

\begin{figure}[H]
    \centering
    \includegraphics[width=1\textwidth]{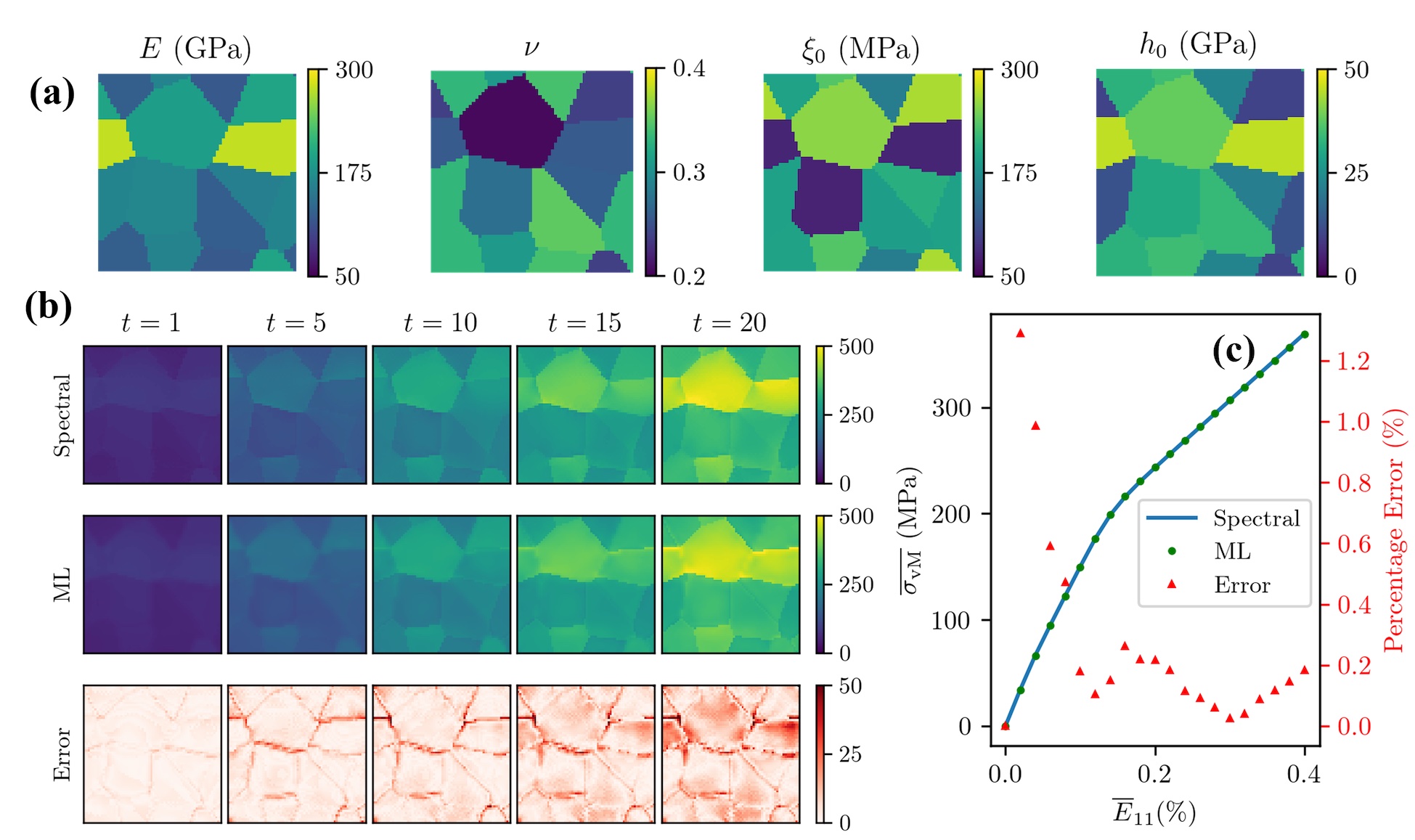}
    \caption{\textbf{(a)} Material properties for a microstructure patch consisting of 10 grains. \textbf{(b)} The local von Mises stress fields based on the solution obtained by the spectral method (above) and the ML model (middle). Also, the absolute error between the ML and the spectral method is plotted (bottom), Units: MPa. \textbf{(c)} The mean von Mises stress and the error (in percentage) between the ML model and the spectral method vs. the strain, the normal component of the mean Green-Lagrange strain, $\bar{E}_{11}$.}
    \label{fig:10grains}
\end{figure}

Fig. \ref{fig:10grains} (a) depicts the chosen material properties for the test case of the 10-grain microstructure. Fig. \ref{fig:10grains} (b) shows the stress fields for the simulations obtained via the spectral method and the ML model at different load steps. Moreover, the absolute error between the ML model and the spectral method is depicted. The stress fields are well predicted with a maximum absolute error of around 50 MPa. Fig. \ref{fig:10grains} (c) depicts the curve of the mean von Mises stress versus the applied strain. The result shows that the mean value of the von Mises stress is accurately approximated during tensile loading with a maximum error of $1.2 \%$. It should be noted that this error is slightly larger than the error observed for the case of the 5-grain microstructure due to the larger length of the grain boundaries in this patch as mentioned above.

\begin{figure}[H]
    \centering
    \includegraphics[width=1\textwidth]{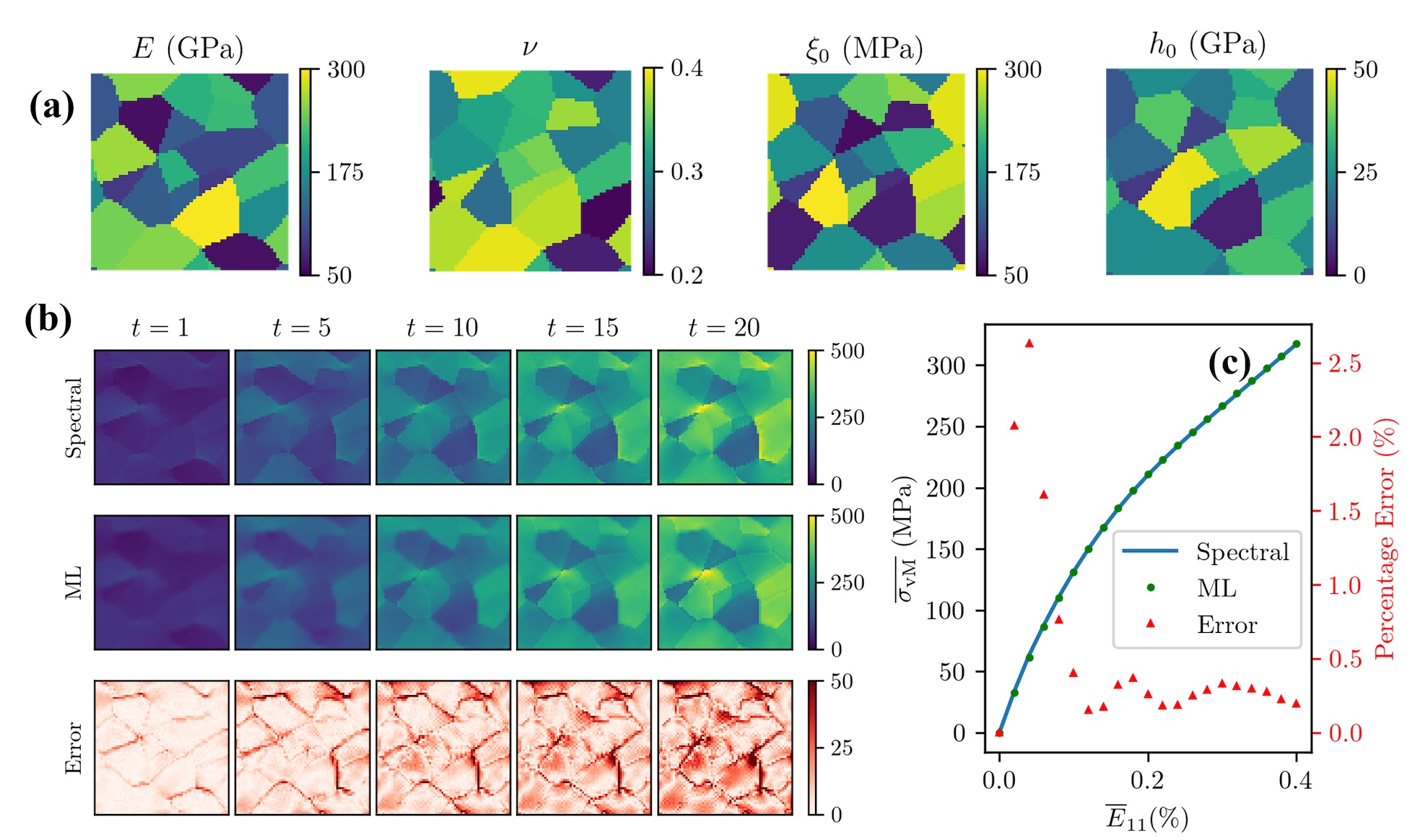}
    \caption{\textbf{(a)} Material properties for a microstructure patch consisting of 20 grains. \textbf{(b)} The local von Mises stress fields based on the solution obtained by the spectral method (above) and the ML model (middle). Also, the absolute error between the ML and the spectral method is plotted (bottom), Units: MPa. \textbf{(c)} The mean von Mises stress and the error (in percentage) between the ML model and the spectral method vs. the strain, the normal component of the mean Green-Lagrange strain, $\bar{E}_{11}$.}
    \label{fig:20grains}
\end{figure}

In the third example, we test the ML model for 20-grain structures. Fig. \ref{fig:20grains} (a) depicts the chosen material properties, which are randomly selected. Fig. \ref{fig:20grains} (b) illustrates the von Mises stress fields at different load steps for the spectral method and ML-based prediction. It can be seen in the figure that the stress fields calculated by the two methods are quite similar. To investigate this finding in more detail, the absolute error is plotted in Fig. \ref{fig:20grains} (b) (bottom row). Here, more pixels have an error of around $50$ MPa than the geometries with less number of grains, see Fig. \ref{fig:5grains} (b) (bottom row) and \ref{fig:10grains} (b) (bottom row). The error increases significantly around the grain boundaries due to the size of the grain, or insufficient numerical resolution. This example is a great case to discuss the issue of grain size, which influences the prediction errors. For the case of structures within 20 grains, some grains are relatively smaller than the case of 10-grain structures; hence, a finer resolution will be needed. Fig. \ref{fig:20grains} (c) shows the stress-strain curve for the spectral method and the ML model together with the error between these two curves. As shown in this plot, the error in average von Mises stress is up to 2.5\%, which is larger than the case of 5- and 10-grain structures, see Fig. \ref{fig:5grains} (c) and Fig. \ref{fig:10grains} (c). This is expected as the total length of the grain boundaries (regions with higher error in stress prediction) is increased compared to the cases with model microstructures based on larger grain sizes discussed above. 

\subsection{Model microstructures with varying material properties}
\begin{figure}[H]
\includegraphics[width=1\textwidth]{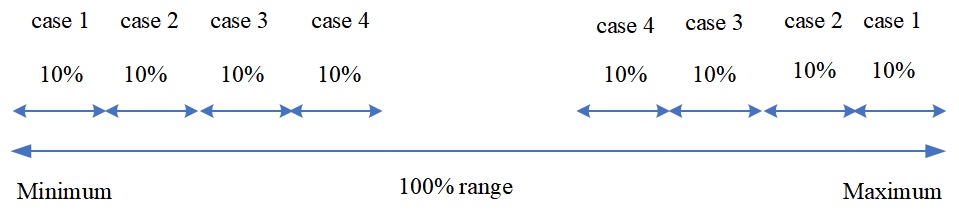}
\caption{The schematic illustration of the material properties chosen at random for four cases. Minimum and maximum values are reported in Table \ref{tab:range_properties}.}
\label{fig:range_properties}
\end{figure}

In this subsection, we investigate the range of material properties considered to predict stress fields. For the training dataset, the whole range ($100\%$ range of the material properties that are reported in Table \ref{tab:range_properties}) is considered for the random selection of the material properties ($E$, $\nu$, $h_0$, $\xi_0$). Four different cases are investigated herein to examine the effect of the range of material properties. For instance, in Case 1, the material properties are randomly selected in two segments - (a) $10\%$ length of the range near to the minimum, (b) $10\%$ length of the range near to the maximum. The material properties of half of the grains and the rest of the grains are randomly selected in segments (a) and (b), respectively. We shift these two $10 \%$ segments toward the middle of the range for the subsequent cases. Fig. \ref{fig:range_properties} schematically depicts how the material properties are selected for four cases. 
This analysis is performed to study the effect of the material property contrast on the quality of the stress field prediction. 

For each of the four material property ranges shown in Fig. \ref{fig:range_properties}, 50 polycrystals, containing 10 grains, are randomly created via the Voronoi tessellation technique. Subsequently, the ML model is evaluated, and the MAE among all 50 geometries is computed for these four cases and is shown in Table \ref{tab:MAE_range:material}. According to this table, the MAE decreases by shifting the segments of the range of the material properties toward the center. It should be noted that the contrast ratio of the material properties in Case 1 is greater than in Cases 2, 3, and 4. In other words, the larger contrast ratio results in larger errors in stress field prediction. In fact, a shorter range of material properties; for example, $80\%$ of the range, is more reliable than the whole range. This result indicates that a wider range for a training dataset can be considered, i.e., $80\%$ of the range can be used for an ML prediction.

\begin{table}[h]
		\centering
		\caption{The mean absolute error of the test dataset for different ranges of material properties shown in Fig. \ref{fig:range_properties}.}
		\label{tab:MAE_range:material}
		\begin{tabular}{|c|c|c|c|c|}
			\hline
			Case & 1  & 2 & 3 & 4 \\ \hline
			MAE (MPa) & 3.662 & 1.925 & 1.519 & 1.265 \\
			\hline
		\end{tabular}
	\end{table}

\begin{figure}[H]
    \centering
    \includegraphics[width=1\textwidth]{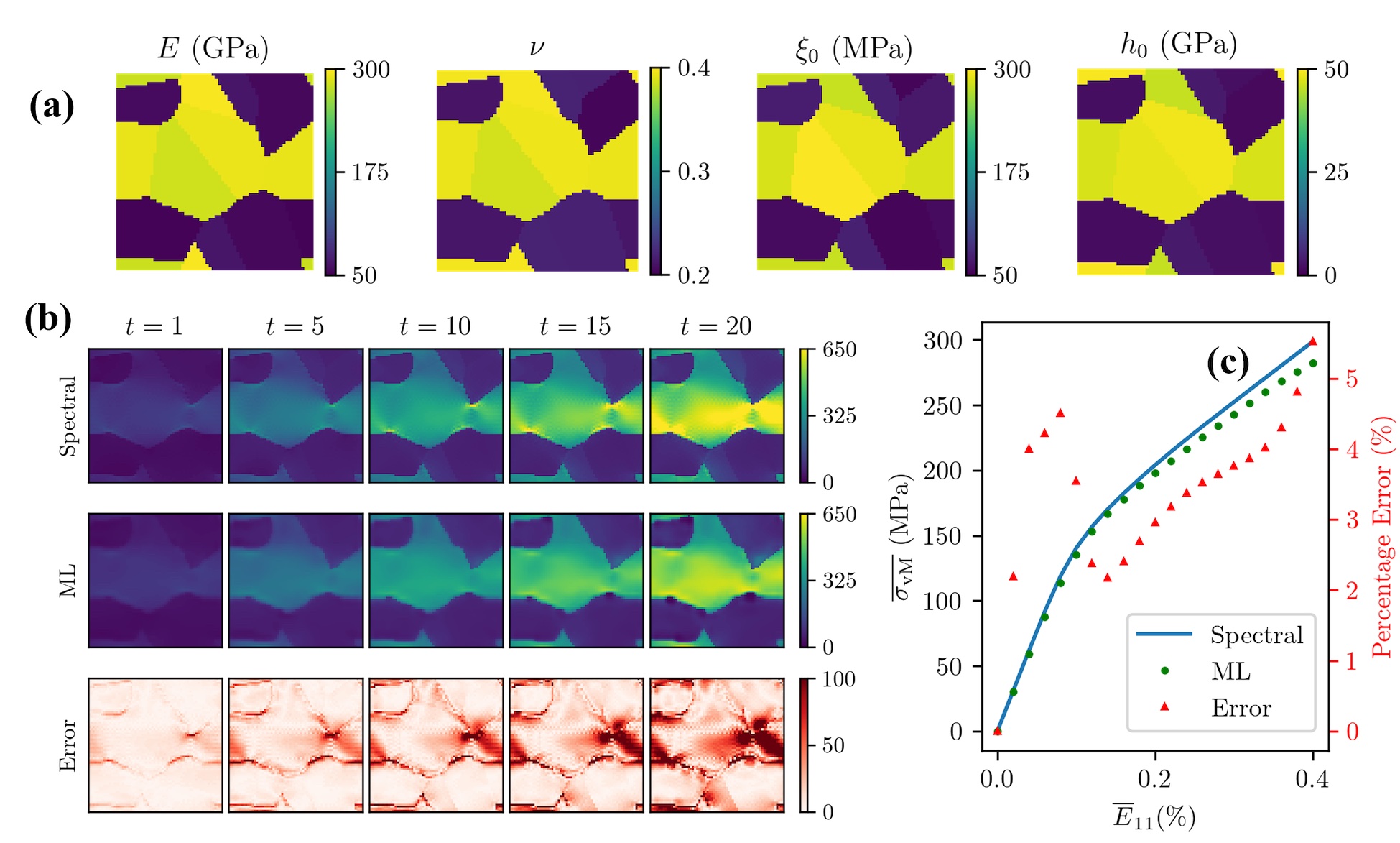}
    \caption{\textbf{(a)} Material properties for a microstructure patch consisting of 10 grains in Case 1. \textbf{(b)} The local von Mises stress fields based on the solution obtained by the spectral method (above) and the ML model (middle). Also, the absolute error between the ML and the spectral method is plotted (bottom), Units: MPa. \textbf{(c)} The mean von Mises stress and the error (in percentage) between the ML model and the spectral method vs. the strain, the normal component of the mean Green-Lagrange strain, $\bar{E}_{11}$.}
    \label{fig:10grains_case1}
\end{figure}

Next, two examples, one geometry in Case 1, associated with a higher contrast ratio, and the other geometry in Case 3 (lower contrast ratio) are examined to illustrate the effect of the range of material properties on the ML prediction. Fig. \ref{fig:10grains_case1} (a) depicts the selected material properties in case 1. The figure shows that the contrast is significantly larger than in other cases in the training dataset. For instance, Young's modulus ($E$) is around $50$ GPa for half of the grains, while it is around $300$ GPa for the rest of the grains. Likewise, the high contrast values exist in the other material properties ($\nu$, $h_0$, and $\xi_0$). Fig. \ref{fig:10grains_case1} (b) depicts the snapshots of the stress fields for the spectral method, the ML model, and the absolute error between these two methods. As shown in the figure, the error is relatively larger than the cases in which material properties are randomly selected in the entire range (cases discussed in the previous section). Here, some of the pixels have an absolute error of around $100$ MPa. In Fig. \ref{fig:10grains_case1} (c), the stress-strain curve and the relative error of these two methods are plotted. As shown in the figure, the percentage error of the ML prediction for the mean von Mises stress is approximately $5\%$, which is almost twice the maximum error in the previous section.    

\begin{figure}[H]
    \centering
    \includegraphics[width=1\textwidth]{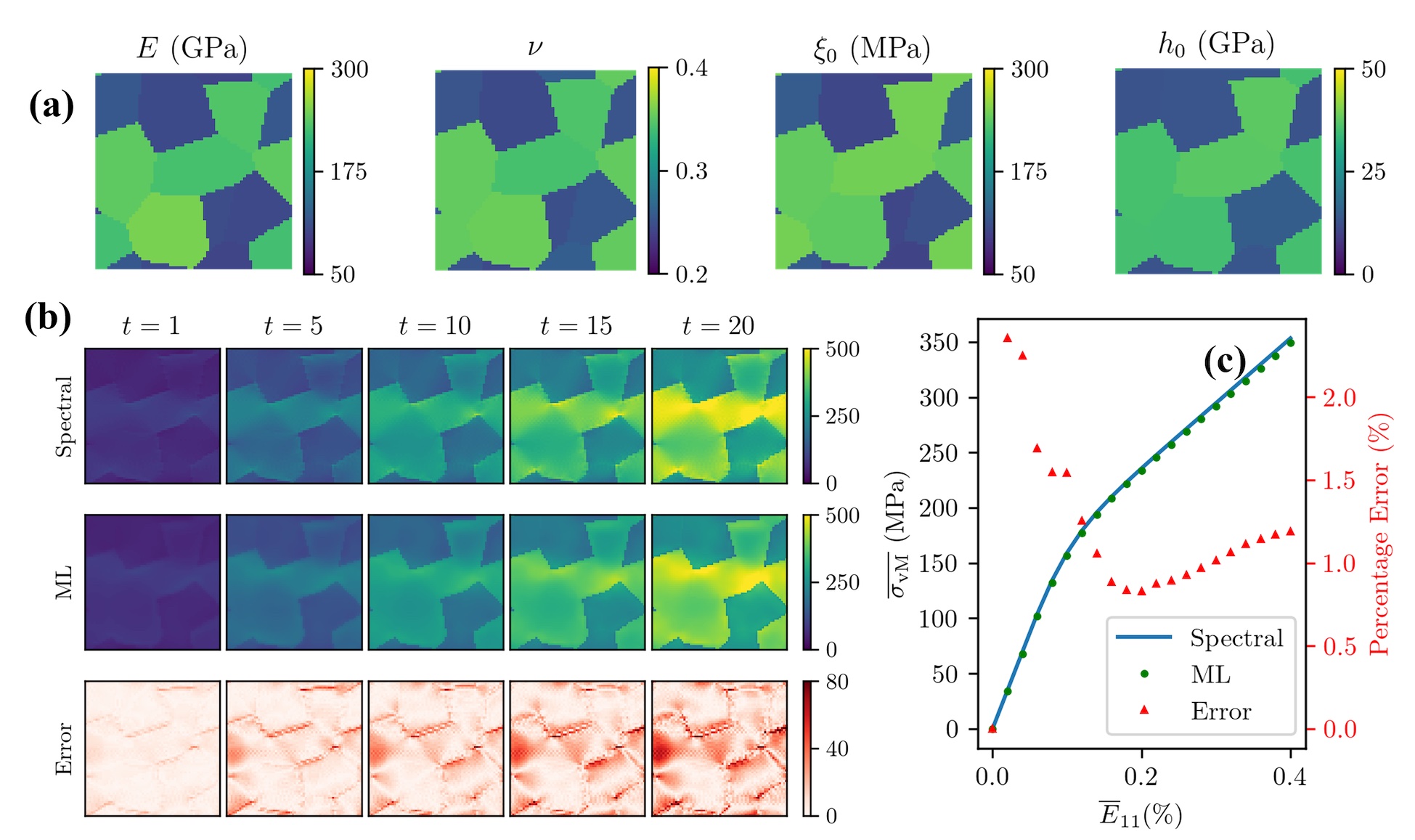}
    \caption{\textbf{(a)} Material properties for a microstructure patch consisting of 10 grains in Case 3. \textbf{(b)} The local von Mises stress fields based on the solution obtained by the spectral method (above) and the ML model (middle). Also, the absolute error between the ML and the spectral method is plotted (bottom), Units: MPa. \textbf{(c)} The mean von Mises stress and the error (in percentage) between the ML model and the spectral method vs. the strain, the normal component of the mean Green-Lagrange strain, $\bar{E}_{11}$.}
    \label{fig:10grains_case3}
\end{figure}

Similarly, among all 50 geometries in Case 3, one geometry is selected to compare the results with Case 1. Fig. \ref{fig:10grains_case3} (a) depicts the chosen material properties of a polycrystal in case 3. In Fig. \ref{fig:10grains_case3} (b), the von Mises stress fields for the methods of spectral and ML are shown at different load steps. Moreover, the absolute error between these two methods is plotted in Fig. \ref{fig:10grains_case3} (b) (bottom). As shown in the figure, the error of some pixels is about $80$ MPa, which is smaller compared to the error in Case 1, see Fig. \ref{fig:10grains_case1} (b) (bottom). Fig. \ref{fig:10grains_case3} (c) depicts the stress-strain curve for the spectral method and ML model. Also, the error between these two methods is plotted. The figure indicates that the prediction of the ML model in Case 3 (with a maximum error of $2 \%$) is more accurate than Case 1 in which the maximum error is around $5 \%$ as shown in Fig. \ref{fig:10grains_case1} (c). 

\subsection{Microstructure morphologies not in the training dataset}
\begin{figure}[H]
    \centering
    \includegraphics[width=0.92\textwidth]{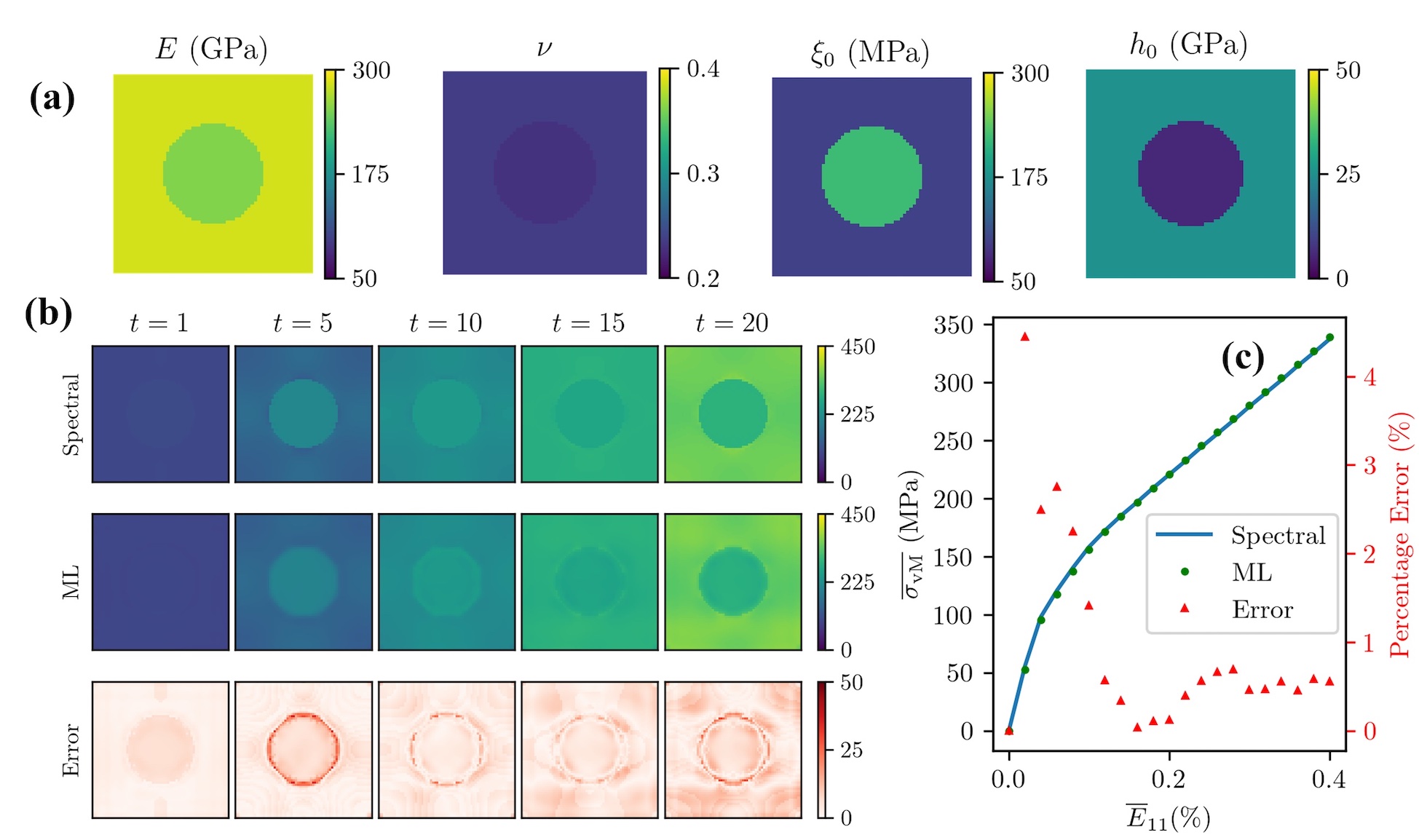}
    \caption{\textbf{(a)} Material properties for a microstructure patch containing a circular inclusion. \textbf{(b)} The local von Mises stress fields based on the solution obtained by the spectral method (above) and the ML model (middle). Also, the absolute error between the ML and the spectral method is plotted (bottom), Units: MPa. \textbf{(c)} The mean von Mises stress and the error (in percentage) between the ML model and the spectral method vs. the strain, the normal component of the mean Green-Lagrange strain, $\bar{E}_{11}$.}
    \label{fig:2grains_circle}
\end{figure}

In this subsection, we explore the ML model for microstructures which are geometrically far from the ones included in the training dataset. The domain contains only two regions in these examples, namely, matrix and inclusion. Two different cases of circular and square-shaped inclusions are investigated. For each case, 50 test case studies are generated, in which the geometries are fixed while the material properties are randomly selected in the whole range of Table \ref{tab:range_properties}. Fig. \ref{fig:2grains_circle} illustrates the viscoplastic mechanical response of the matrix-inclusion for the circular inclusion. The chosen material properties are shown in Fig. \ref{fig:2grains_circle} (a). The von Mises stress fields based on the spectral method and the ML model as well as the error between these methods are depicted in Fig. \ref{fig:2grains_circle} (b). The figure indicates that the error (about $50$ MPa) is comparable to the error in the stress prediction for the microstructures in the training dataset (10-grain polycrystals). The stress-strain curve for this case is also plotted in Fig. \ref{fig:2grains_circle} (c) for the spectral method and the ML model. Furthermore, the error between these two approaches is plotted in this curve. Interestingly, the mean von Mises stress is predicted with a maximum error of $4 \%$; however, most of the load steps have errors below $1 \%$ as shown in this plot. 

\begin{figure}[H]
    \centering
    \includegraphics[width=0.92\textwidth]{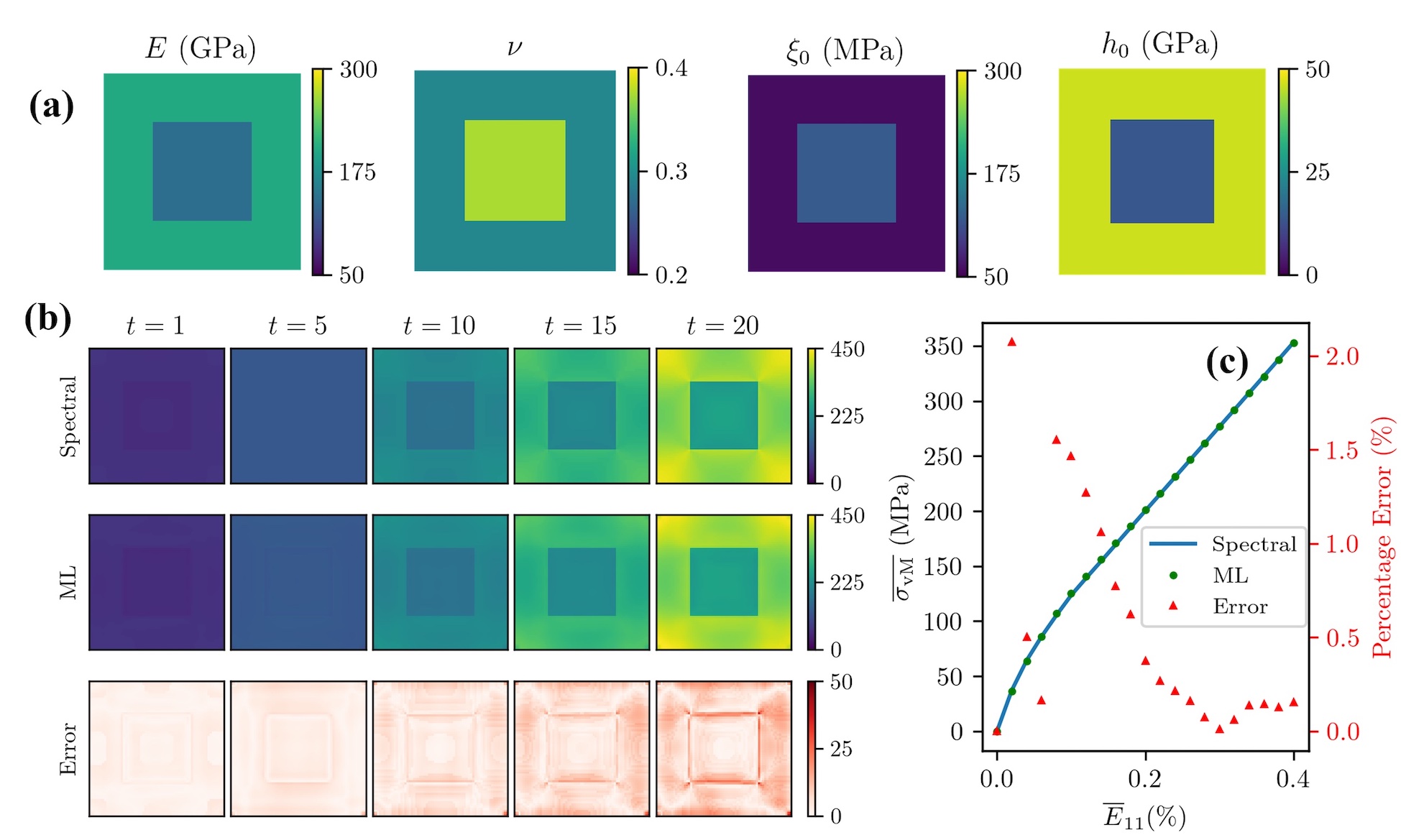}
    \caption{\textbf{(a)} Material properties for a microstructure patch containing a square inclusion. \textbf{(b)} The local von Mises stress fields based on the solution obtained by the spectral method (above) and the ML model (middle). Also, the absolute error between the ML and the spectral method is plotted (bottom), Units: MPa. \textbf{(c)} The mean von Mises stress and the error (in percentage) between the ML model and the spectral method vs. the strain, the normal component of the mean Green-Lagrange strain, $\bar{E}_{11}$.}
    \label{fig:2grains_square}
\end{figure}

Analogously, the ML model is evaluated for square inclusion. Fig. \ref{fig:2grains_square} (a) depicts the selected material properties for the case of square inclusion. In Fig. \ref{fig:2grains_square} (b), the von Mises stress fields for the spectral method and the ML model are plotted. The results show that besides errors at the interfaces, the relative most significant error, approximately $50$ MPa, occurs at the corners, which is challenging in numerical methods as well as in the ML prediction. Furthermore, the curve of mean von Mises stress versus the mean applied strain is plotted for these methods in Fig. \ref{fig:2grains_square} (c). In this figure, the percentage error between the spectral method and the ML model is plotted. As shown here, the mean value of the von Mises stress is estimated with the maximum error of $2\%$. 

The MAE among all 50 tests of the circular inclusion is $1.42$ MPa, which is slightly lower than the MAE of $1.58$ MPa in the case of the square inclusion. This comparison shows that the smoother shapes lead to a slightly better prediction. Again, the more significant errors happen around the sharp interfaces between matrix and inclusion. Additionally, the error increases as the loading continues due to the accumulative error of the prediction at each load step.

\subsection{Loading histories not in the training dataset}
\begin{figure}[H]
    \centering
    \includegraphics[width=1\textwidth]{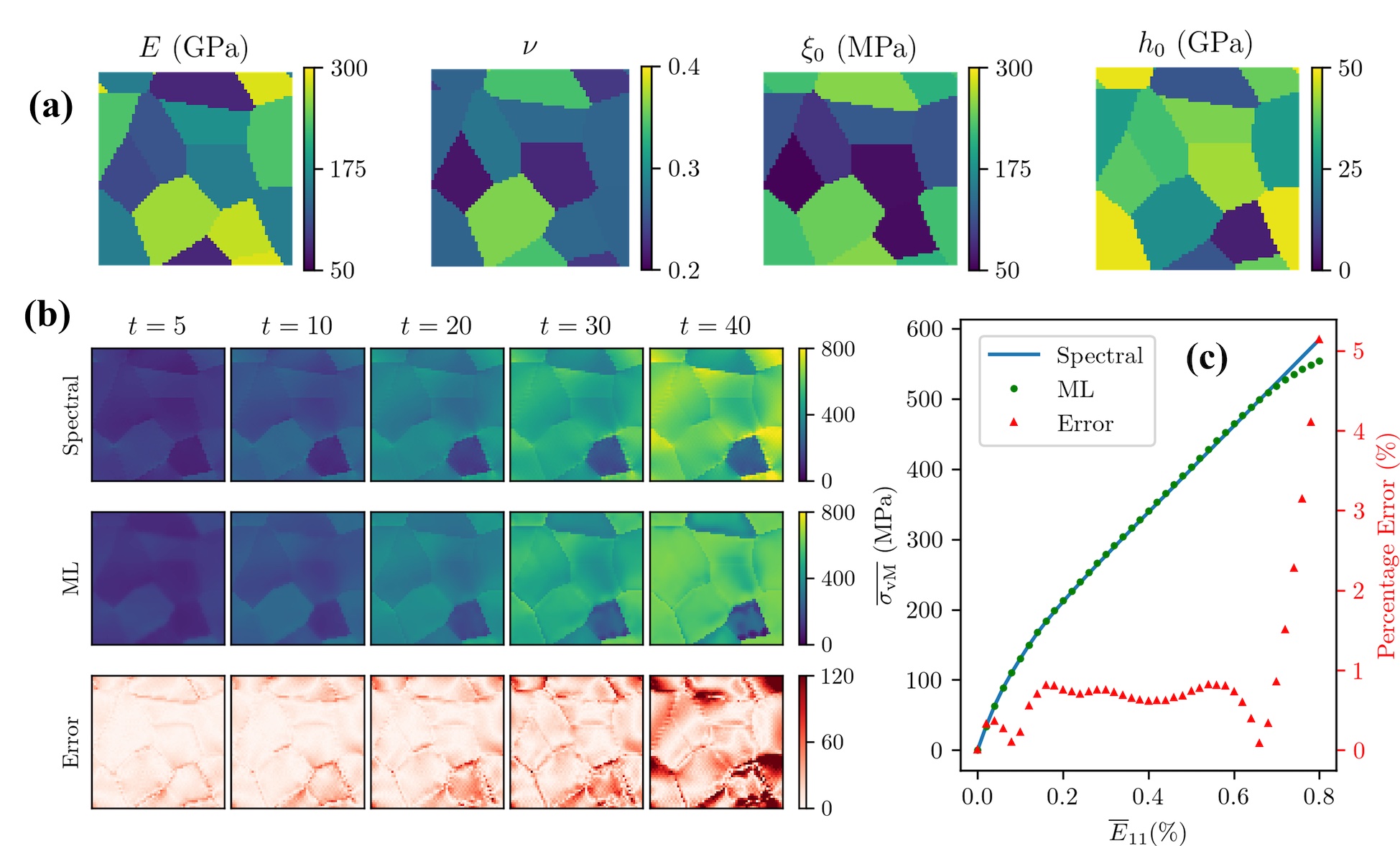}
    \caption{\textbf{(a)} Material properties for a microstructure patch consisting of 10 grains for the case of the loading beyond the training dataset. \textbf{(b)} The local von Mises stress fields based on the solution obtained by the spectral method (above) and the ML model (middle). Also, the absolute error between the ML and the spectral method is plotted (bottom), Units: MPa. \textbf{(c)} The mean von Mises stress and the error (in percentage) between the ML model and the spectral method vs. the strain, the normal component of the mean Green-Lagrange strain, $\bar{E}_{11}$.}
    \label{fig:10grains_furtherLoading}
\end{figure}

This subsection investigates the ML model for the tensile loading beyond the dataset. Although the training data is based on the loading up to $\bar{E}_{11}=0.4\%$, the ML model can be used for further loading, from $\bar{E}_{11}=0.4 \%$ to $\bar{E}_{11}=0.8\%$ provided that the $\sigma_{\mathrm{vM}}$ does not exceed the maximum $\sigma_{\mathrm{vM}}$ in the training dataset. In our training data, the maximum von Mises stress is $852$ MPa. However, the von Mises stress of more than $500$ MPa reduces the prediction quality due to the insufficient data at those stress levels in the training process. To confirm this, we refer to Fig. \ref{fig:hist}, where the histogram profile of the von Mises stress at the final step of the loading, i.e., increment $t=20$, is plotted. According to Fig. \ref{fig:hist}, only 0.084\% of the pixels in the whole training dataset have a von Mises stress level above $500$ MPa. Fig. \ref{fig:10grains_furtherLoading} (a) indicates the chosen material properties for the case of the loading beyond the dataset. Fig. \ref{fig:10grains_furtherLoading} (b) depicts the von Mises stress fields at various load steps. As shown in the figure, the error increases further after $t=20$, or $\bar{E}_{11}=0.4 \%$. In particular, when the stress is larger than $500$ MPa, the prediction deteriorates. Fig. \ref{fig:10grains_furtherLoading} (c) shows the stress-strain curve for further loading. According to this curve, the error notably increases after $\bar{E}_{11}=0.8\%$, where the stress is greater than $500$ MPa. However, the additional loading beyond the level in the training (i.e. $\bar{E}_{11}=0.4\%$ to $\bar{E}_{11}=0.7\%$) still produces good quality results with MAE below 1.0\%, until the box stress reaches about 500 MPa at around $\bar{E}_{11}=0.7\%$.

\subsection{Computational efficiency}

Here, we compare the computational efficiency of the spectral method and the ML model. For a fair comparison of the computational time, we consider a single core of an Intel\textsuperscript{\textregistered} Core\texttrademark i9900K clocked at $3.60$ GHz. For solving the mechanical BVPs using the spectral method, we utilize the DAMASK package \cite[][]{roters2019damask}. Since the spectral method needs a number of iterations to reach the converged solution, this numerical method is computationally expensive. On the other hand, the machine learning model can predict the stress fields without any iteration. In order to compare the efficiency, we do the simulation for 50 different geometries containing 10 grains under tensile loading with 20 increments. The average run time of DAMASK to obtain a solution for one simulation is approximately 75 seconds (averaged over 50 simulations). In contrast, the run time of the ML model for one case is about 0.15 seconds (averaged over 50 cases). Hence, in our comparison, the speed-up is about a factor of 500. In other words, the ML model is approximately 500 times faster than the spectral method to obtain the solution for all increments of the loading. It should be noted that the speed-up significantly depends on the resolution. If we increase the number of grid points, the computational time of the spectral method increases significantly. The number of iterations also depends on the load level, the nonlinearity of the system, and the complexity of the geometry. The reported speed-up above is meant only as a rough estimate. 

\section{Conclusions and Outlooks}
\label{sec:conclusions}

Any improvement to increase computational efficiency within reasonable accuracy would be great in the mechanical analysis of polycrystals or composites. In this work, we have developed a machine learning (ML) model for the viscoplastic response of polycrystals or composites. The ML model can accurately and fast predict the history-dependent von Mises stress in the J2 plasticity framework. Compared with the conventional numerical approaches such as spectral methods to solve the relevant mechanical boundary value problems, our ML model estimates the solution relatively fast with acceptable accuracy. The main reason for this speed-up is that spectral methods require a number of iterations to reach the converged solution, while the operations in the ML model are of a fixed number and require only one forward pass. The main advantage of this surrogate model is that the ML model can be obtained with a few configurations of material properties. In this work, we have generated $1000$ different geometries with random selections of material properties for each grain. Although the training data is based on 10-grain polycrystals, the obtained ML model can be used for a wider range of grain numbers and shapes. It should be noted that the number of combinations of these cases is uncountable. However, we showed that only $1000$ geometries, which are randomly generated, can be sufficient to obtain a surrogate model for predicting the local stress fields. 

Of course, if we include more geometries with more combinations of material configurations and various geometries, the ML model can be improved. We highlight the main key point of our work: the surrogate model can rapidly predict the local fields such as von Mises stress without a significant drop in accuracy. In comparison with the spectral method, the ML model is nearly 500 times faster for the resolution of $64 \times 64$. Indeed, if we increase the resolution, the computational time of the spectral method will be increased more than the ML model due to the nature of the iterative procedure in numerical approaches such as the spectral method. In other words, the speed-up of the ML is much more significant for the higher resolutions. Moreover, our ML model is notably precise in the prediction of the mean value of the von Mises stress fields, which can be used in homogenization techniques.

Despite the numerous advantages of our ML model, it is restricted to the specified loading and boundary conditions (BCs). In this work, the obtained ML model is based on one increment of tensile loading under periodic boundary conditions. Indeed, the model is not general for arbitrary loading or BCs. As a perspective for future works, this model can be extended for predicting local fields such as all components of strain and stress fields under arbitrary BCs. Furthermore, our approach can be combined with a physics-informed neural network, which has been vastly focused on in recent years and will be our focus in the future.       
    
%\bibliographystyle{unsrtnat}
%\bibliography{references.bib} 

\end{document}